# Analysis of the main factors for the configuration of green ports in Colombia

Abraham Londoño-Pineda[1,*], Tatiana Arias-Naranjo[2], Jose Alejandro Cano[3]

[1]Universidad de Medellín, Faculty of Economic and Administrative Sciences, Medellín, Colombia; [2]Universidad de Medellín, Medellín, Colombia; [3]Universidad de Medellín, Medellín, Colombia; [*]corresponding author email: alondono@udem.edu.co address: Carrera 87 # 30-65, Medellín, Colombia.

**Abstract**

This study analyzes the factors affecting the configuration and consolidation of green ports in Colombia. For this purpose, a case study of the maritime cargo ports of Cartagena, Barranquilla and Santa Marta is performed addressing semi-structured interviews to identify the factors contributing to the consolidation of green ports and the factors guiding the sustainability management in the ports that have not yet been certified as green ports. The results show that environmental regulations are a starting point but are not the key factor to consolidate as a green port. Factors such as investment in clean technologies, integration with interest groups, and evaluation and monitoring of relevant environmental indicators have become the differential factors allowing Santa Marta's port to have been certified and recognized as a green port. As a conclusion, the conversion of Colombian ports to green ports should not be limited to the attainment of certifications, such as EcoPort certification, but should ensure the contribution to sustainable development through economic, social, and environmental dimensions, environmental, and through the achievement of the SDGs.

**Introduction**

A green port can be defined as a port performing its activities causing minimal damage to the environment and society, providing improvement measures and quality control of air, water, noise, and waste (Chen et al., 2019; Fahdi, Elkhechafi, & Hachimi, 2019; Zis, 2019). This aligns the green ports initiative with the international political agenda on sustainable development, contributing to the achievement of several sustainable development goals (SDGs), expressed in the UN 2030 Agenda (Londoño & Baena, 2017; Londoño & Cruz, 2019; Londoño, 2015).

The green port designation is given through a certification granted by the European Sea Ports Organization (ESPO) which in turn promotes the Port Environmental Review Systems (PERS) standard, which is the only environmental management standard specific to the port sector, whose compliance is certified through the firm Lloyd's Register Quality Assurance (EcoPorts, 2019). Among the advantages generated by EcoPorts is the promotion of safe trade and efficiency in port operations, the entry into the logistics chain of companies that only invest and make businesses with certified ports, the attraction of new clients, minimization of adverse impacts on the community and the environment, and the improvement in the corporate image in society (Akgul, 2017; Satir & Doğan-Sağlamtimur, 2018).

Colombia was the first country outside the European Union having an EcoPort, which was granted to the Santa Marta's port for the first time in 2013, mainly due to the transportation of coal, because the mining activity can not be conceived without having an obligation and responsibility with the community and the environment (Moore, 2008; Zeng, Liu, He, Ma, & Wu, 2018). This achievement is due, to a large extent, to the international entrepreneurship capabilities of the management team of the Port Society of Santa Marta related to customer orientation, proactivity, and the value creation for stakeholders (Causado & Londoño, 2015; Tabares, Alvarez, & Urbano, 2015), which have contributed to reflect in their mission and vision the commitment to sustainability.

Likewise, it should be noticed that the environmental policies have a direct impact on the environmental regulation established by the ports (Tsai et al., 2018; Woo, Moon, & Lam, 2018), and these policies must explain the sanctions, tax incentives, and rewards for those ports that promote environmentally sustainable practices (Sheu, Hu, & Lin, 2013). Therefore, it is required a balance between the demands of the ports towards the maritime operators and the incentives that allow such operators to maintain profitability, ensuring harmonization between environmental regulation and economic leverage (Tseng & Pilcher, 2019; Wan, Zhang, Yan, & Yang, 2018).

**Case Study**

In this research, we followed an integrated multiple case study (Yin, 2013), since the analysis units are the three main ports of the Colombian Atlantic coast (Cartagena, Barranquilla and Santa Marta), and the variables to be analyzed are institutional, associated with the environmental policy and the attitude of the seaport management team towards the entrepreneurship. Information of the study was collected through a semi-structured interview applied to strategic managers of the ports, whose questions are presented in Table 1.

Table 1. Semi-structured interview questions applied to port managers

| Factors | Questions |
| --- | --- |
| Legislative factors | What is the impact of the environmental policy and environmental legislation in Colombia in achieving the EcoPort certification? |
| | How environmental policy and environmental legislation in Colombia affect the configuration of a green port? |
| Internal factors | How have the factors associated with the capacity of the management team influenced the obtaining of the EcoPort certification? |
| | How do the factors associated with the capacity of the management team contribute towards the configuration of a green port? |
| Certification EcoPort | Does the port have EcoPort certification? |

In the case studies addressed in this investigation, the relevance of the cases prevails instead of a criterion of statistical representativeness (Gibbert & Ruigrok, 2010; Yin, 2013), which is associated with the importance of the selected analysis units. Thus, the selected ports are the most important of the Atlantic Coast in Colombia in terms of infrastructure, transit, the volume of goods, and storage capacity.

**Results**

The following is a synthesis of the responses obtained for the ports of Cartagena, Barranquilla, and Santa Marta.

**Port of Cartagena**

Environmental policy and environmental legislation serve as a guide for ports to be environmentally sustainable. This port complies with the legal environmental obligations and moves towards environmental management and this is corroborated by the decreases in diesel and energy consumption, which results in lower $CO_2$ emissions per year. Even the port of Cartagena has been nominated by the ESPO as the port facility that has most safeguarded and contributed to nature and ecosystem in its influence area. Therefore, this port is not far from becoming a green port, since there are environmental practices that even go beyond compliance with the law, however, it is necessary to follow the phases of this new initiative, and especially involve the different stakeholders in this process.

**Port of Barranquilla**

The legal obligations indicated in the environmental policies are a starting point taken by the port, however, it requires environmental management focusing on the configuration of a green port. Although the port of Barranquilla is not certified as a green port, it performs actions aimed at environmental management such as having an air emissions permit, granted by the local environmental authority, which implies conducting annual air quality studies. Besides, vehicles entering the port must have current technical and mechanical certifications. Similarly, the port has liquid dumping permits, has a solid waste management program that is disseminated to the entire business group and contractors that enter the port facilities. Different environmental aspects have been intervened in the port, but high investments in clean technologies are required to achieve an effective transformation towards a green port.

**Port of Santa Marta**

The environmental legislation is the starting point for a green port certification and the environmental planning process is coordinated with the National Environmental System (SINA), in addition, decree 3083 of 2007 of Colombia provides some guidelines regarding the direct coal loading in ports exporting this mineral, however, there is no specific legislation in Colombia regarding green ports, as the only environmental management standard specific to the port sector (PERS) is promoted by ESPO. However, environmental legislation in Colombia promotes good environmental practices, which results in obtaining this type of certification. Some of the factors that are evaluated, monitored and have been crucial for the port of Santa Marta to obtain an EcoPort certification are: the efficiency in the use of water for port processes, initiatives such as direct loading of coal, air quality control, replacement of gasoline equipment with electrical equipment, having and protecting the last coral relic over the bay of Santa Marta at the operating docks, as well as the social work accomplished by the Santa Marta Port Society Foundation. Therefore, the green ports certification is not limited in complying legislative requirements, but also requires financial investment for environmental management, and especially requires the integrative work of the related groups affected by the operation of the port and the continuous training of the port staff.

In summary, the port of Cartagena is focused on the integration of the stakeholders, the port of Barranquilla is focused on high investments in clean technologies, and the port of Santa Marta highlights that the factors contributing to obtain the EcoPort certification are related to technology investment, orientation towards stakeholders, training of port staff in sustainability, and strict monitoring of environmental indicators.

**Conclusions**

In the ports studied, the importance of environmental regulations and policies is highlighted to create an institutional framework that promotes environmental protection and sustainability, however, if there is no commitment and proactive attitude of the management team of port societies the regulatory framework will be insufficient. Because there is no specific policy for green ports in Colombia, it is necessary to focus on more important factors such as the integration of the stakeholders, investments in clean technologies, training of port staff in sustainability, and monitoring of environmental indicators; these factors allowed some ports to be certified as green ports and receive nominations and permits demonstrating good environmental practices.

**References**

Akgul, B. (2017). Green Port / Eco Port Project-Applications and Procedures in Turkey. In F. A. M. Y. I. M. M. K. O. Drusa M. Torok I. (Ed.), *IOP Conference Series: Earth and Environmental Science* (Vol. 95). Institute of Physics Publishing.

Causado, L. M., & Londoño, A. A. (2015). Proposal for a new Born Global or International New Venture taxonomy [Propuesta para una nueva clasificación de Born Global (BG) o nuevos Emprendimientos Internacionales (INVs)]. *Espacios*, *36*(15), 2.

Chen, J., Zheng, T., Garg, A., Xu, L., Li, S., & Fei, Y. (2019). Alternative Maritime Power application as a green port strategy: Barriers in China. *Journal of Cleaner Production*, *213*, 825–837.

EcoPorts. (2019). EcoPorts. Retrieved August 6, 2019, from https://www.ecoports.com/

Fahdi, S., Elkhechafi, M., & Hachimi, H. (2019). Green Port in Blue Ocean: Optimization of Energy in Asian Ports. In B. B. M. K. H. H. Kaicer M. Addaim A. (Ed.), *2019 International Conference on Optimization and Applications, ICOA 2019*. Institute of Electrical and Electronics Engineers Inc.

Gibbert, M., & Ruigrok, W. (2010). The "What" and "How" of Case Study Rigor: Three Strategies Based on Published Work. *Organizational Research Methods*, *13*(4), 710–737.

Londoño, A. (2015). El enfoque de gobernanza en la evaluación del desarrollo sostenible a escala local (caso del departamento de Antioquia, Colombia). *Revista Mexicana de Ciencias Agrícolas*, *1*, 257–236.

Londoño, A., & Baena, J. J. (2017). Análisis de la relación entre los subsidios al sector energético y algunas variables vinculantes en el desarrollo sostenible en México en el periodo 2004-2010. *Gestión Y Política Pública*, *26*(2), 491–526.

Londoño, A., & Cruz, J. G. (2019). Evaluation of sustainable development in the sub-regions of Antioquia (Colombia) using multi-criteria composite indices: A tool for prioritizing public investment at the subnational level. *Environmental Development*, (April), 1–22.

Moore, P. (2008). Colombia's coal king. *Mining Magazine*, *199*(3), 28+30–31.

Satir, T., & Doğan-Sağlamtimur, N. (2018). The protection of marine aquatic life: Green Port (EcoPort) model inspired by Green Port concept in selected ports from Turkey, Europe and the USA. *Periodicals of Engineering and Natural Sciences*, *6*(1), 120–129.

Sheu, J.-B., Hu, T.-L., & Lin, S.-R. (2013). The key factors of green port in sustainable development. *Pakistan Journal of Statistics*, *29*(5), 755–767.

Tabares, A., Alvarez, C., & Urbano, D. (2015). Born globals from the resource-based theory: A case study in Colombia. *Journal of Technology Management and Innovation*, *10*(2), 154–165.

Tsai, Y.-T., Liang, C.-J., Huang, K.-H., Hung, K.-H., Jheng, C.-W., & Liang, J.-J. (2018). Self-management of greenhouse gas and air pollutant emissions in Taichung Port, Taiwan. *Transportation Research Part D: Transport and Environment*, *63*, 576–587.

Tseng, P. H., & Pilcher, N. (2019). Evaluating the key factors of green port policies in Taiwan through quantitative and qualitative approaches. *Transport Policy*, (December 2018), 1–11.

Vega De La Cruz, L. O., & Ortiz Pérez, A. (2018). Procesos más relevantes del control interno de una empresa hotelera. Semestre Económico, 20 (45), 217-231. https://doi.org/10.22395/seec.v20n45a8

Wan, C., Zhang, D., Yan, X., & Yang, Z. (2018). A novel model for the quantitative evaluation of green port development – A case study of major ports in China. *Transportation Research Part D: Transport and Environment*, *61*, 431–443.

Woo, J.-K., Moon, D. S. H., & Lam, J. S. L. (2018). The impact of environmental policy on ports and the associated economic opportunities. *Transportation Research Part A: Policy and Practice*, *110*, 234–242.

Yin, R. K. (2013). *Case Study Research: Design and Methods (Applied Social Research Methods)* (Fifth). London: SAGE Publications.

Zeng, X., Liu, Z., He, C., Ma, Q., & Wu, J. (2018). Quantifying surface coal-mining patterns to promote regional sustainability in Ordos, Inner Mongolia. *Sustainability (Switzerland)*, *10*(4).

Zis, T. P. V. (2019). Green ports. In H. N. Psaraftis (Ed.), *Sustainable Shipping: A Cross-Disciplinary View* (pp. 407–432). Springer International Publishing.